\documentclass{osa-article}

\journal{oe}

\usepackage{amsmath,amsfonts,amssymb}
\usepackage{graphicx}
\usepackage{caption}
\usepackage{subfigure}
\usepackage[subfigure]{tocloft}
\usepackage{setspace}
\usepackage{tocloft}
\usepackage{gensymb}
\usepackage{xfrac}
\usepackage{tikz}
\usepackage{wasysym}
\usepackage[ruled,vlined]{algorithm2e}
\usepackage{pdfpages}
\usepackage{epstopdf}

\newcommand{\Pb}{\textnormal{Pb}}
\newcommand{\Zr}{\textnormal{Zr}}
\newcommand{\Ti}{\textnormal{Ti}}
\newcommand{\Nb}{\textnormal{Nb}}
\newcommand{\OO}{\textnormal{O}}
\newcommand{\PM}{\textnormal{pm}}
\newcommand{\V}{\textnormal{V}}
\newcommand{\dd}{\textnormal{d}}
\newcommand{\Gg}{\textnormal{g}}

\cftpagenumbersoff{figure}
\cftpagenumbersoff{table} 

\begin{document}

\title{High pixel number deformable mirror concept utilizing piezoelectric hysteresis for stable shape configurations}

\author{R. Huisman,\authormark{1,*} M. Bruijn,\authormark{1} S. Damerio,\authormark{2} M. Eggens,\authormark{1} S.N.R. Kazmi,\authormark{1} A.E.M. Schmerbauch,\authormark{3} H. Smit,\authormark{1} M.A. Vasquez-Beltran,\authormark{3} E. van der Veer,\authormark{2} M. Acuautla,\authormark{3} B. Jayawardhana,\authormark{3} and B. Noheda\authormark{2}}

\address{\authormark{1}SRON Netherlands Institute for Space Research, Landleven 12, 9747AD Groningen, The Netherlands\\
\authormark{2}Zernike Institute for Advanced Materials, University of Groningen, Nijenborgh 4, 9747AG Groningen, The Netherlands\\
\authormark{3}Engineering and Technology Institute of Groningen, University of Groningen, Nijenborgh 4, 9747AG Groningen, The Netherlands}

\email{\authormark{*}R.Huisman@SRON.nl}

\begin{abstract}
We present the conceptual design and initial development of the Hysteretic Deformable Mirror (HDM). The HDM is a completely new approach to the design and operation of deformable mirrors for wavefront correction in advanced imaging systems. The key technology breakthrough is the application of highly hysteretic piezoelectric material in combination with a simple electrode layout to efficiently define single actuator pixels. The set-and-forget nature of the HDM, which is based on the large remnant deformation of the newly developed piezo material, facilitates the use of time division multiplexing (TDM) to address the single pixels without the need for high update frequencies to avoid pixel drift. This, in combination with the simple electrode layout, paves the way for upscaling to extremely high pixel numbers ($\geq 128\times 128$) and pixel density ($100/mm^2$) deformable mirrors (DMs), which is of great importance for high spatial frequency wavefront correction in some of the most advanced imaging systems in the world.
\end{abstract}

\section{Introduction}
\label{sect:intro} 
The search for planets outside our solar system (so called exoplanets), experienced a great boost after the discovery of the first exoplanets in the 1990's. Since then already more than $4000$ exoplanets have been found with mostly indirect detection techniques, such as the Transit Method and the Radial Velocity Method \cite{Wright2012}. These techniques do not collect the light from the very faint planet itself, but observe the effect it has on its accompanying star. Although these techniques are very suitable to discover exoplanets, they do not provide information about the chemical composition of a possible atmosphere surrounding the planet, and thus about the possibility of life existing on this planet. To obtain detailed information on the composition of planet atmospheres, it is necessary to directly observe the light from the planet.

To be able to directly detect the planet, the extreme contrast between the star light and that of the accompanying planet ($10^{-10}$ for an Earth-like planet around a Sun-like star in the visible wavelength regime) requires the application of a coronagraph, a starshade or even a space interferometer, to avoid the star from completely drowning out the light from the planet. Future large space telescopes, dedicated to direct imaging of exoplanets like LUVOIR \cite{LUVOIR2019} and HABEX \cite{Habex2019}, will rely on deformable mirrors (DMs) to achieve clean and stable wavefronts over long periods of time, while reducing the negative effect of star light speckles originating from small surface shape errors of the optics. For the DM in such an instrument this translates to: $< 1nm$ wavefront error correction, shape stability $\leq 10 pm RMS$ over $10s$ of minutes, high spatial frequency correction ($128\times 128$ pixels) and extreme reliability.

Currently state-of-the-art DMs are developed by Boston Micromachines Corporation (BMC) and AOA Xinetics (AOX).
Both have working DMs with $64\times 64$ pixels. BMC uses MEMS as single electrostatic actuators \cite{Bifano2011}. They utilize
a multiplexer fed by a single high voltage (HV) amplifier to address the different pixels, but the design requires separate wiring for every individual pixel. Another critical issue with this design is that of shape instability as a result of pixel
drift. AOX develops both surface-normal and surface-parallel piezoelectric DMs based on lead magnesium niobate (PMN) \cite{Wirth2013}. The AOX DMs are very stable but the discrete nature of the PMN actuators again requires separate wiring for every pixel.
It is also difficult to reduce the actuator pitch, which makes application of high pixel number DMs in small imaging systems more difficult. Two $48\times 48$ pixel AOX DMs are currently baselined for NASA's Wide-Field InfraRed Survey Telescope (WFIRST), scheduled to be launched in the mid 2020s \cite{Spergel2015}.
Other noticeable DM developments, which focus on space applications, are the $44$ pixel DM from the University of M\"{u}nster \cite{Rausch2016}, the monomorph DM from CILAS \cite{Cousty2016}, the $57$ pixel DM from TNO \cite{Kuiper2016} and the $24$ pixel MADRAS DM \cite{Laslandes2017}. An interesting development, making use of radiation pressure as actuation mechanism to address high pixel numbers in a dense configuration is described in \cite{Riaud2012}. Microscale Inc. specifically addresses the problem of extensive wiring. They have developed lead zirconate titanate (PZT) based DMs with special ASIC boards to minimize the pixel wiring and reduce the mass of the total system \cite{Prada2017}. Finally, a detailed overview of different DM techniques for adaptive optics is provided in \cite{Madec2012}.

Scaling of the current state-of-the-art DMs to the needed capabilities for a direct exoplanet imaging space mission is not trivial, in particular concerning the demand on pixel number and pixel density and practical operability (wire bonding, harness and electronics) in a light weight and reliable design. This paper presents the conceptual design of a Hysteretic Deformable Mirror (HDM), which utilizes large hysteresis in the piezo actuators in combination with an elegant pixel layout to overcome typical problems like pixel drift and fabrication and reliability of harness and electronics when considering extremely high pixel number and pixel density DMs.
We present the principle of operation, the conceptual design and the current baseline for the fabrication of the hardware. We discuss the development of the unique hysteretic piezo material, which is of paramount importance for the functioning of the HDM, and we advance the strategy to control the shape of the mirror surface.

\section{HDM concept}\label{HDM_Concept}

\subsection{Principle of operation}\label{HDM_Principle}
\begin{figure}[htbp]\centering
\begin{tikzpicture}
\node at (-4,0) {\includegraphics[scale=0.15]{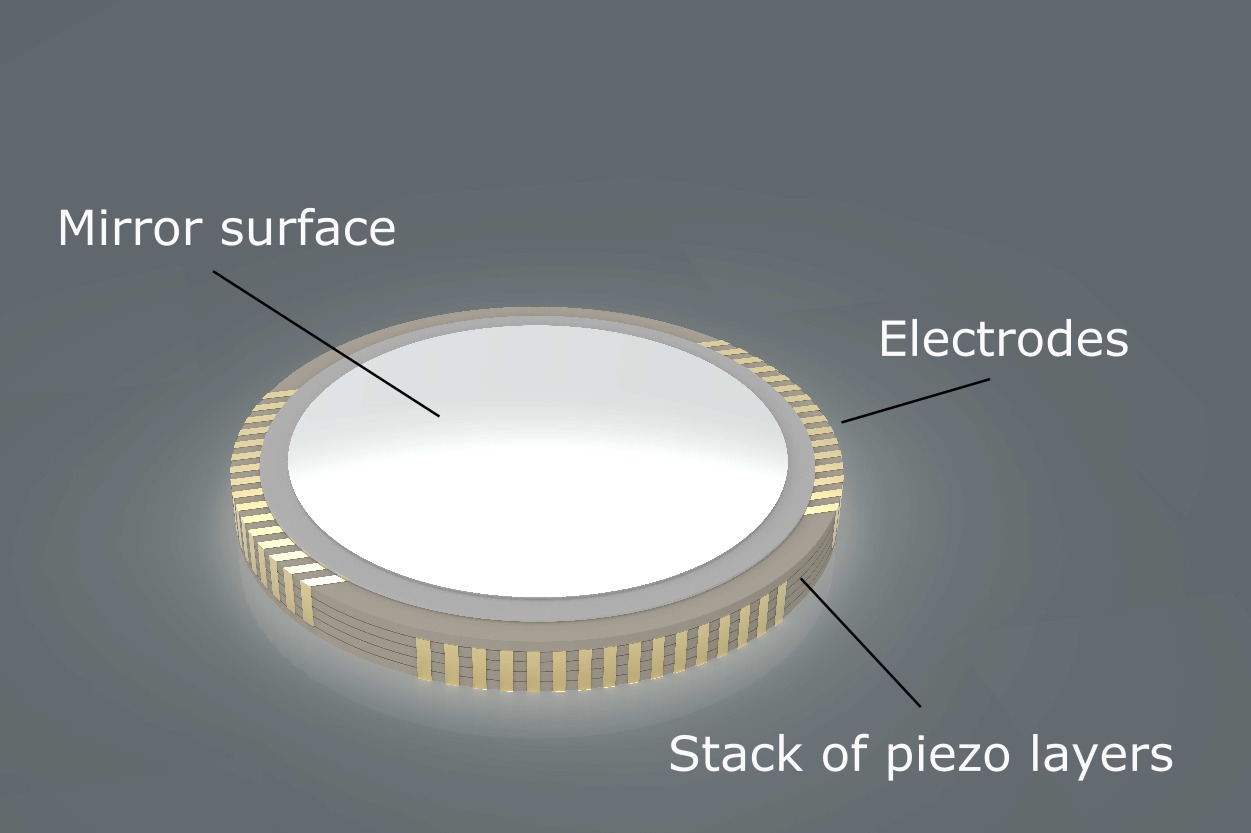}};
\node at ( 2.8,0) {\includegraphics[scale=0.131]{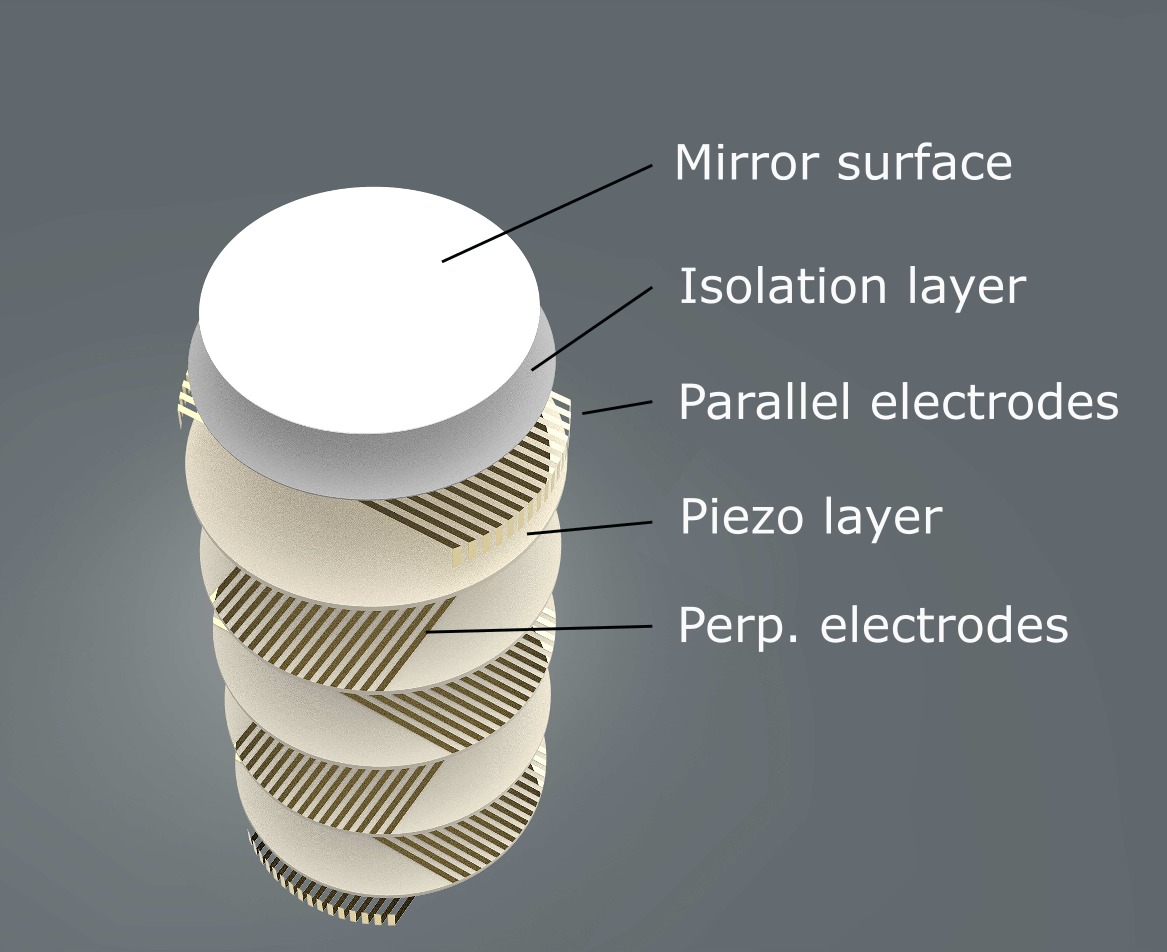}};
\node at (-4,-2.5) {a)};
\node at ( 2.8,-2.5) {b)};
\end{tikzpicture}
\caption{a) HDM conceptual design. b) Exploded view. Deformation of the mirror is realized by a stack of piezo layers each with perpendicular top and bottom electrodes. Electrode strips which are directly above each other are electrically connected and commanded by a single input voltage. For clarity, only 5 layers with $16\times 16$ pixels are shown, the final hardware however will consist out of a larger number of these layers with $128\times 128$ pixels to improve the performance of the DM.} 
\label{fig:HDMconcept}
\end{figure}

Figure \ref{fig:HDMconcept} presents the conceptual design of the HDM. Deformation of the reflective top layer is realized by a stack of thin piezo layers separated from each other by top and bottom electrodes made of arrays of parallel metallic strips. The strips of the bottom and top electrodes are oriented perpendicular to each other and they are equally aligned for all layers, in such a way that the bottom electrode strips of the different layers lay directly above each other and are electrically connected, the same being true for the top electrode strips.
Single actuator pixels are simply defined by this electrode layout. The electric field, generated when applying a voltage over any two perpendicular strips, causes local deformation of the sandwiched piezo layer at the crossing of those particular electrodes. The simulations which support this idea are presented in Section \ref{Section FEM}.

For clarity, only a very limited number of layers are shown in Figure \ref{fig:HDMconcept}. As for standard piezo stack actuators, the required voltage to reach a certain total deformation, decreases when more layers with intermediate electrodes are applied. 
The required number of layers depends on the final material properties and thickness of the piezo layers (as discussed in Section \ref{Material}), and the required stroke for the considered application. This is discussed in more detail in Section \ref{Discussion}.

The application of low voltages (in the order of $10s$ of Volts) makes it possible to scale down the electrode width and pitch. It also eliminates the need for a HV-amplifier. Another advantage is that because of the large ratio between pixel pitch and layer thickness, the electric field (E-field) is constrained to a small surface array, limiting the cross talk between the pixels and again facilitating a very compact design.

\begin{figure}
\begin{center}
\begin{tabular}{c}
\includegraphics[height=5cm]{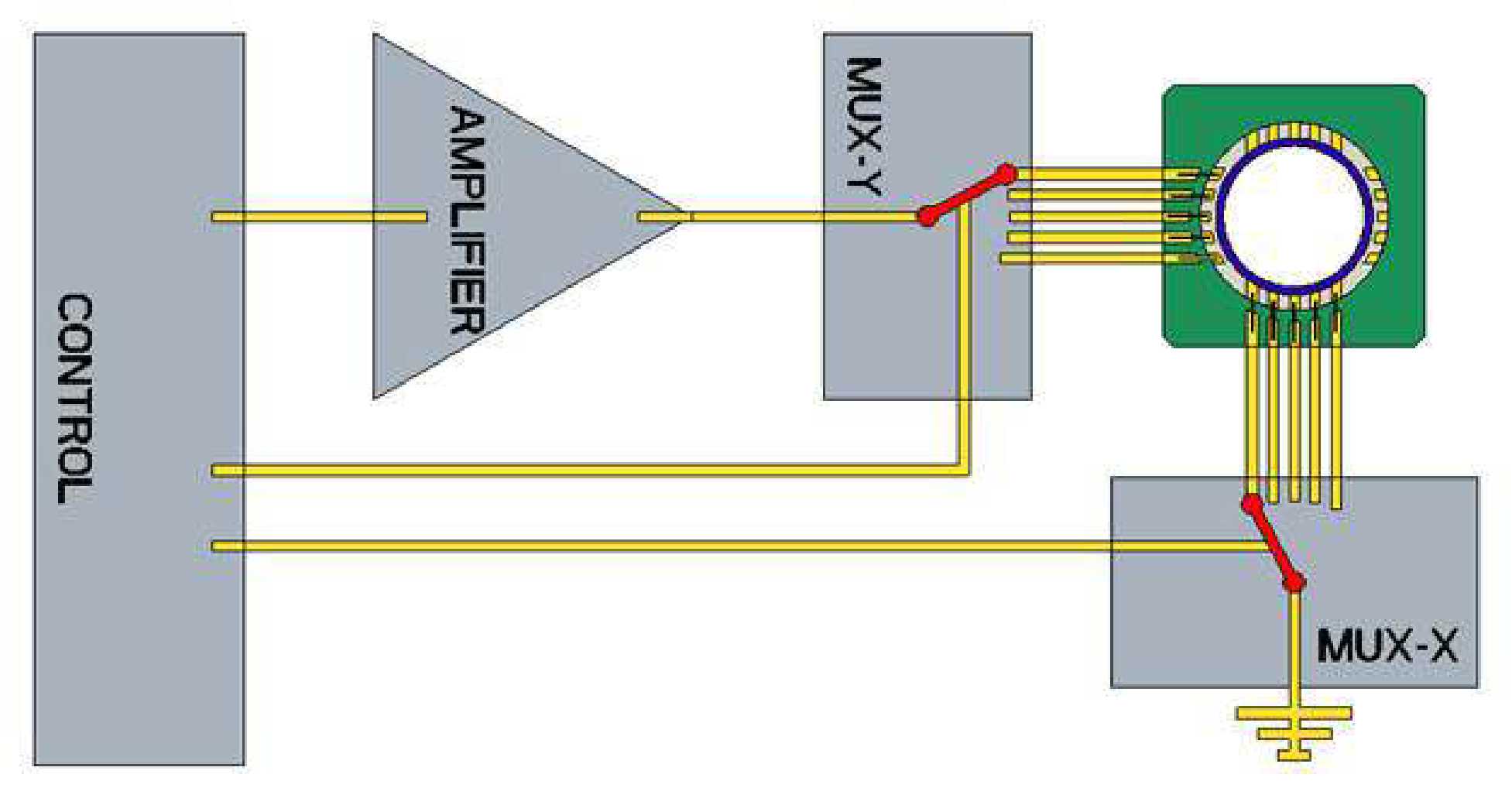}
\end{tabular}
\end{center}
\caption 
{\label{fig:HDMsystem}
HDM system layout. Both the amplifier and the MUXs are set by the control system. All vertically stacked electrodes are interconnected and are addressed by a single input from the amplifier. MUX-X is connected to ground. For clarity, only 5$\times$5 electrodes are used in this example.} 
\end{figure}

Figure \ref{fig:HDMsystem} shows how the single pixels are addressed by the electronics. The use of strip electrodes minimizes the harness to a single wire per electrode. Therefore, the harness scales with the square root of the number of pixels $2\sqrt{n}$, rather than $n$. This reduces the number of separate wires for a $128\times 128$ pixel DM from 16384 to 256. We apply time division multiplexing to address the single pixels, which means that we only need a single amplifier instead of $n$ of them.

The use of strip electrodes to define individual pixels can only work when the actuated pixels keep their commanded position after completely removing the locally generated E-field. We solved this fundamental problem by developing piezo material with an asymmetric/distorted butterfly loop, which functions as a set-and-forget actuator. We have achieved this by means of Niobium doping of piezoelectric PZT, which we call PNZT, as described in more detail in Section \ref{Material}.

It is also this highly nonlinear characteristic of the PNZT material which guarantees that, although during actuation of a single pixel shape changes are observed at all crossings along the addressed electrodes (see Section \ref{Section FEM}), remnant deformation will only be observed at the commanded pixel location. The details about how this is exploited in the control of the system are provided in Section \ref{Controlstrat}.

The set-and-forget nature and the application of low voltage input signals, greatly reduces the power dissipation of the DM, because no constant HV-input or high update frequency is required to avoid pixel drift as a result of charge leakage of the holding capacitor \cite{Horenstein2011} when high pixel number DMs are considered. This is not only advantageous from an operational point of view, but it also limits thermoelastic deformations in the imaging system and eliminates pixel instability as a result of noise in the driver electronics.
The limited number of electronic components and wiring, together with the stiff mechanical design of the HDM, guarantee a lightweight and compact system which is very robust to failure and vibration loads. We consider this to be an important merit of the design, as reliability is by far the most important requirement for space applications.

\subsection{FEA of HDM pixel response} \label{Section FEM}
To support the conceptual idea, a Finite Element Analysis (FEA) has been performed which simulates the electric field generation in the piezoelectric array and the resulting pixel response. As discussed in Section \ref{HDM_Principle}, when an electric potential is placed on one of the top electrode strips, while one of the bottom strips is connected to ground, an electric field is generated in the piezo layers at the crossing of the two strips. This will yield a piezoelectric response and deformation of the respective pixel in all piezo layers.

To simulate the activation of a pixel in a single layer of piezo material, the conceptual structure of the HDM was generated in Comsol Multiphysics 5.4 using the Solid Mechanics and Electrostatics modules, i.e., Piezoelectric Devices multiphysics interface. A thin piezoelectric layer of commercial PZT-5H was created, actuated by 5 parallel strips as top electrodes and 5 parallel strips as bottom electrodes, the latter oriented perpendicularly to the top electrode strips, as previously introduced. We will refer to this configuration as a 5$\times$5 electrode grid. This grid was modelled by adding the thin layer boundary condition to simulate the response of the nanoscale structure and avoid meshing problems.
The outer edges of the model and the bottom boundary were set to be mechanically clamped (modelled by fixed boundary condition). To activate the center pixel, an electric potential of $600V$ was applied to the center top strip while the center bottom strip was connected to ground. The surrounding strips were provided with the floating potential condition. The high voltage was applied to clearly show the deformation over the surface, in practice however, the electric potential for any of the thin piezo layers will be limited to $< 100V$.
The simulation does not include the exotic hysteretic behaviour of the newly developed material, so no remnant deformation is observed. An overview of the used parameters for the model and the simulation are listed in Table \ref{tab:Comsol_parameters}.

\begin{figure}[htbp]\centering
\begin{tikzpicture}
\node at (-3.2,0) {\includegraphics[scale=.3]{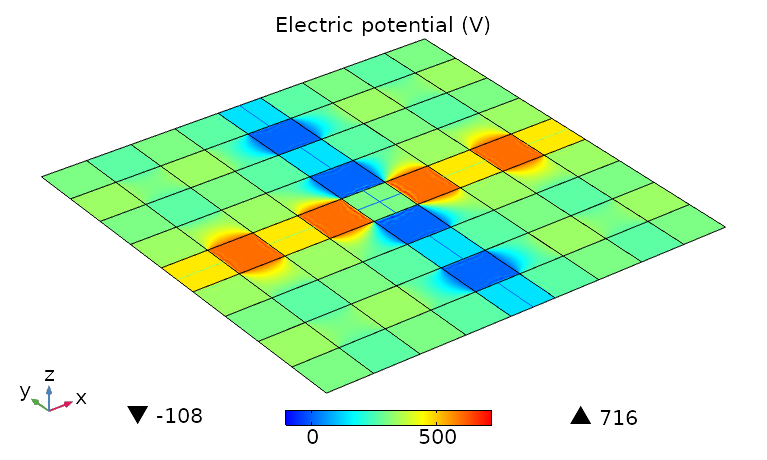}};
\node at ( 3.2,0) {\includegraphics[scale=.3]{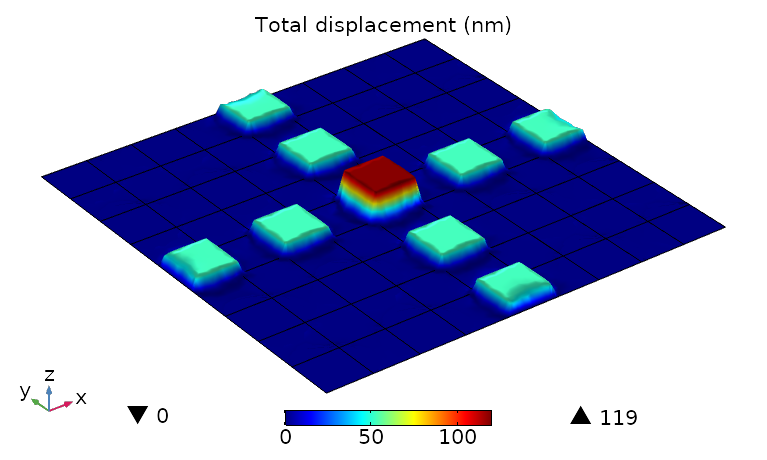}};
\node at (-3.2,-3.5) {a)};
\node at ( 3.2,-3.5) {b)};
\end{tikzpicture}
\caption{FEA of 5$\times $5 pixel array when addressing the center pixel only. An electrical potential of $600 V$ was applied to the mid-top strip while the mid-bottom strip was grounded. The strips were modelled by means of the thin layer boundary condition. a) The distribution of the electric potential in the xy-midplane. b) Deformation of the center and neighboring pixels. The maximum deformation of the center pixel amounts to $119 nm$.} 
\label{fig:Comsol_elP}
\end{figure}

 \begin{table}[ht]
 \caption{Parameters and boundary conditions of the FEA to study HDM pixel addressing.} 
 \label{tab:Comsol_parameters}
 \begin{center}       
 \begin{tabular}{ll} 
 \hline
 \rule[-1ex]{0pt}{3.5ex}  \textbf{Electrodes} &   \\
 \hline\hline
 \rule[-1ex]{0pt}{3.5ex}  Width & 1 mm  \\
 \rule[-1ex]{0pt}{3.5ex}  Pitch & 2 mm   \\
 \hline
 \rule[-1ex]{0pt}{3.5ex}  \textbf{Piezoelectric layer} &    \\
 \hline
 \rule[-1ex]{0pt}{3.5ex}  Material & PZT-5H  \\
 \rule[-1ex]{0pt}{3.5ex}  Thickness & 750 nm  \\
 \hline
 \rule[-1ex]{0pt}{3.5ex}  \textbf{Boundary conditions} &  \\
 \hline
 \rule[-1ex]{0pt}{3.5ex}  Strips & Thin layer boundary condition  \\
 \rule[-1ex]{0pt}{3.5ex}  Center top strip & Grounded  \\
 \rule[-1ex]{0pt}{3.5ex}  Center bottom strip & Electrical potential of 600 V  \\
  \rule[-1ex]{0pt}{3.5ex}  Surrounding strips & Floating potential boundary condition \\
 \hline
 \rule[-1ex]{0pt}{3.5ex}  \textbf{Mesh} &   \\
 \hline 
 \rule[-1ex]{0pt}{3.5ex} Domain elements &  101320  \\
 \rule[-1ex]{0pt}{3.5ex} Boundary elements & 38264  \\
  \rule[-1ex]{0pt}{3.5ex} Edge elements & 4600  \\
  \rule[-1ex]{0pt}{3.5ex} Degrees of freedom solved for & 1271162  \\
 \hline
  \rule[-1ex]{0pt}{3.5ex} \textbf{Computation time} &   \\
 \hline
  \rule[-1ex]{0pt}{3.5ex} Solution time & 8 min 39 sec   \\
 \hline
 \end{tabular}
 \end{center}
 \end{table} 

Figure \ref{fig:Comsol_elP}a shows the distribution of the electric potential when addressing the center pixel only. From this it is clear that not only at the intersection of the active strips, but also at the other pixel locations along both strips, an electric field is generated. Although the field strength is clearly lower at these offset positions, this will result in deformation of the piezo surface at the mentioned offset positions while the potential difference is applied between the two strips, as can be seen in Figure \ref{fig:Comsol_elP}b. 
By using the Preisach operator, and exploiting the properties introduced by the hysteresis, it is possible to determine conditions such that the application of the control signal to every pixel does not modify the remnant deformation of the previously set pixels, therefore, overcoming this potentially limiting effect. This is discussed in more detail in Section \ref{Controlstrat}.

\section{HDM design and manufacturing}\label{Nanotech}

\subsection{HDM design}
\begin{figure}
    \centering
    \includegraphics[scale=0.15]{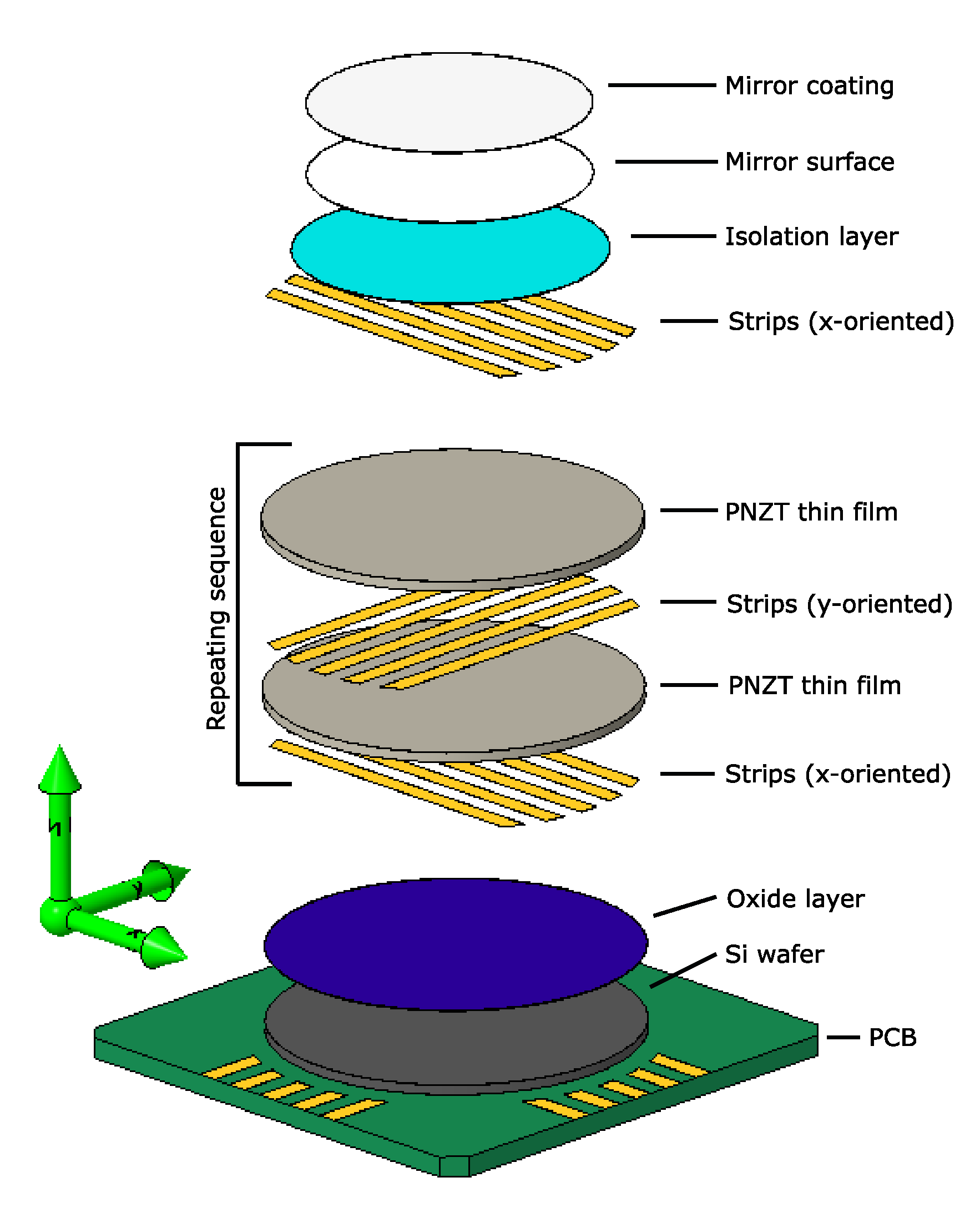}
    \caption{HDM layered structure.}
    \label{fig:HDMlayers}
\end{figure}

Figure \ref{fig:HDMlayers} shows the layered structure of the HDM. The basis consists of a highly resistive Si wafer with a thin layer of thermally grown oxide. On top of this a repeating pattern of the following layers is placed sequentially: an x-axis oriented array of parallel metal strips, a thin piezo layer, an array of y-axis oriented metal strips, and finally another thin piezo layer. The reflective top layer is electrically isolated from the final layer of y-axis oriented strips by a separate isolation layer.

\begin{figure}
    \centering
    \includegraphics[scale=0.5]{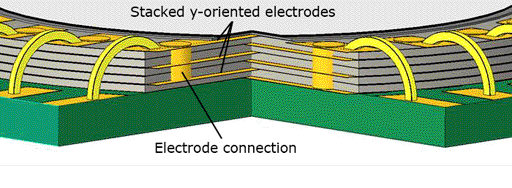}
    \caption{Interconnection of metal strips. Shown is a cut out of the piezo stack (only 5 layers in this example) with intermediate metal strip electrodes. The vertical pillars present the connection of all strips which lie directly above each other.}
    \label{fig:HDMvias}
\end{figure}

As stated earlier, all metal strips which are oriented along the same axis and are positioned directly above each other, are electrically connected. Figure \ref{fig:HDMvias} shows the vias which connect the different strips. The reliable connection of the strips is one of the critical process steps in the realization of the HDM, how this is done is discussed in Section \ref{Manufacturing}. Standard wire bonding technique is used to connect on-chip smaller bond pads to the bigger bond pads on the printed circuit board (PCB) which is attached to the basis.

\subsection{HDM manufacturing}\label{Manufacturing}
\begin{figure}
    \centering
    \includegraphics[scale=0.6]{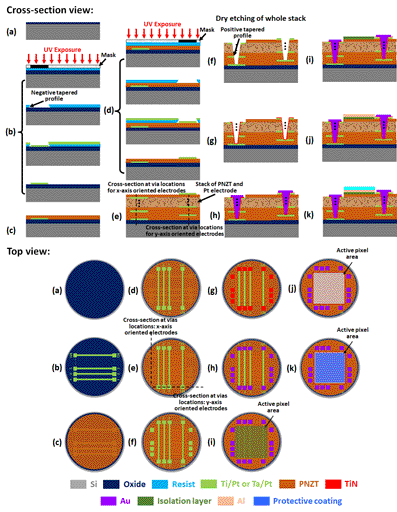}
    \caption{Schematic of the baseline process flow for HDM (cross-section and top view). (a) Carrier wafer with $2 \mu m$ thermally grown oxide. (b) Patterning of x-axis oriented Pt electrode strips on the oxide. (c) Deposition of PNZT thin film by spin coating. (d) Patterning of y-axis oriented Pt electrode strips on the PNZT thin film. (e) Stacking of PNZT thin films between the x-axis and y-axis oriented Pt electrodes. (f) RIE of stack of PNZT thin films sandwiched between the x-axis and y-axis oriented Pt electrode strips. (g) ALD of TiN thin film to ensure interlayer Pt electrode strip connection. (h) Deposition of Au thin film by evaporation or sputtering for on-chip smaller bond pads. (i) Deposition of insulation layer (polymers, oxides or nitride) above the active pixel area to have a smooth and flat surface. (j) Deposition of top reflective layer by lift-off using Al evaporation above the active pixel area. (k) Deposition of protective coating by PECVD of (SiO$_2$/TiO$_2$)  above the mirror surface.}
    \label{fig:fabrication}
\end{figure}

In Figure \ref{fig:fabrication} the current baseline for the fabrication of the HDM is presented. The fabrication starts with a highly resistive Si wafer having $2 \mu m$ thermally grown oxide as depicted in Fig. \ref{fig:fabrication}(a). This oxide layer prevents a possible short between the bottom metal electrode strips on the carrier wafer. Next, the first layer of platinum (Pt) electrodes (x-axis oriented) is patterned onto the thermally grown oxide  by employing the lift-off process, as seen in Fig. \ref{fig:fabrication}(b). For better adhesion of evaporated Pt to the oxide surface, a thin layer ($~5 nm$) of titanium (Ti) or tantalum (Ta) has previously been deposited on the oxide by evaporation. Our preliminary experiments with metal lift-off using image reversal photoresist and metal evaporation resulted in smooth edges of the metal patterns. This in comparison to the rough edges by sputtered metal. That is why we employ thermally evaporated metals for the lift-off process during the electrodes patterning. The thickness of the Pt layer (usually $\geq100 nm$) and choice of adhesion layer (Ti or Ta) will be chosen such that they can withstand the high temperatures applied during the stacking of sol-gel spin-coated PNZT piezo thin films.

A single PNZT thin film is build up out of multiple nano-layers which are sequentially deposited by spin coating. The details of the PNZT thin film deposition process, by the sol-gel method, are provided in Section \ref{PNZThysfilm}. A single PNZT thin film is spin coated on top of the x-axis oriented electrode strips, as shown in Fig. \ref{fig:fabrication}(c). The y-axis oriented Pt strips are then patterned onto the PNZT thin film by the  lift-off process (Fig. \ref{fig:fabrication}(d)). A sequence of PNZT thin film spin coating, patterning of x-axis oriented electrodes, PNZT thin film spin coating, and patterning of y-axis oriented electrodes is repeated to have a stack of PNZT thin films and Pt electrodes as shown in Fig \ref{fig:fabrication}(e). The alignment of the x-axis and y-axis oriented Pt electrode strips is done using the alignment markers patterned at the backside of the Si wafer. These back side alignment markers provide a positional accuracy of $1-2 \mu m$ during the stacking process. The deposition of the PNZT thin film on the metal strips might results in an uneven topography that needs to be smoothened by means of mechanical planarization, either after each PNZT thin film deposition or once the whole stacking process is completed. This will depend on the evolved topography.

The stacked electrode strips are interconnected by either dry or wet chemical etching of the vias. These vias can either be etched into the entire PNZT thin films and x-axis/y-axis oriented Pt electrodes stack as shown in Fig. \ref{fig:fabrication}(f), or by etching of PNZT thin films in successive steps before the next x-axis/y-axis oriented Pt electrodes are deposited. The etch rate of PZT using the dry etching technique is slow ($~100 nm/min$) \cite{Bale2001}. This is expected to be similar for PNZT. Therefore, dry etching techniques like reactive ion etching (RIE) or focused ion beam etching (FIB) are limited to prototype production or applications which require a limited number of PNZT layers. Positive sloped tapered vias made by RIE, as those in Fig. \ref{fig:fabrication}(f), facilitate the interconnection of x-axis/y-axis oriented Pt electrodes by metal deposition. For a thick stack of PNZT thin films and electrodes ($>100 \mu m$), Atomic Layer Deposition (ALD) can be employed due to its exceptional conformality on high-aspect ratio structures to ascertain interlayer connection of electrodes. The etched vias are then conformally covered (see Fig. \ref{fig:fabrication}(g)) with thin titanium nitride (TiN) film (typically few nanometer thick) by ALD to connect the Pt electrodes in the stack. Note that the TiN film by ALD covers the entire wafer, therefore TiN film needs to be removed from the wafer by a wet etching solution (ammonium hydroxide, hydrogen peroxide, and DI water), while covering the vias locations with photoresist. Afterwards, small on-chip gold (Au) bond pads, as depicted in Fig. \ref{fig:fabrication}(h), are deposited either by evaporation or sputtering at vias locations. These small bond pads are then connected with the bigger bond pads on the PCB using the wire bonding technique. The wire bonding is the last step in the fabrication line once the deposition of isolation layer, reflective top layer, and protective coating is carried out above the active pixel area, Fig. \ref{fig:fabrication}(i-k). Laser ablation is being ruled out, because this technique is quite rough and has a lot of redeposition, which might not result in clean vias with proper exposed side access to the metal lines.

\begin{figure}
    \centering
    \includegraphics[scale=0.7]{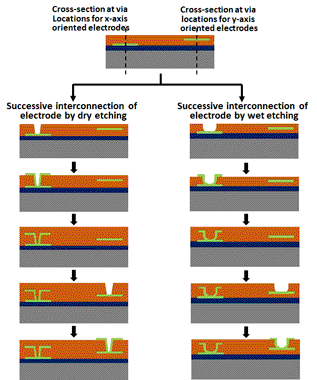}
    \caption{Schematic of the process flow for successive interconnection of x-axis/y-axis oriented Pt electrode strips by dry and wet chemical etching.}
    \label{fig:wedging}
\end{figure}

An alternate approach to interconnect the x-axis/y-axis oriented electrode strips at similar positions is show in Figure \ref{fig:wedging}. Once the first set of x-axis and y-axis oriented Pt electrodes is patterned, the thin film of PNZT is spin coated above the y-axis oriented electrodes. Next, the vias are etched by removing PNZT above the small metal pads of the x-axis oriented electrodes, away from the active pixel area, by RIE (positive tapered profile) or wet etching (isotropic profile). Another set of x-axis oriented Pt electrodes is patterned by evaporation, connecting the second set of x-axis oriented electrodes with the underlying x-axis oriented electrodes. The same process can be repeated to connect the y-axis oriented electrode strips. This cycle can be repeated until the last set of x-axis/y-axis oriented electrodes is connected with the underlying electrodes.

Because of its faster etch rate ($~1.54 \mu m/min$) \cite{Wang2019}, wet chemical etching of PZT is quite attractive compared to dry etching of PZT. For our application, we require a wet etchant selective towards the mask layer (typically photo resist) that protects the active pixel area and Pt electrodes. The wet etching is advantageous as multiple wafers with PNZT and electrodes can be etched simultaneously and can be bonded together to have a thicker PNZT and metal electrode stack.

The choice of isolation layer, directly beneath the mirror surface, is primarily based on its ease of processing, low residual stress, adhesion to the underlying interfaces of PNZT/Pt, and its electrical and mechanical properties. It serves to prevent a short with the mirror surface besides providing a flat and smooth surface for top reflective layer deposition. We are currently performing short loop experiments to choose either spin coated polymers (PDMS, Polyimide (LTC9305) etc.) or Plasma-enhanced chemical vapor deposition (PECVD) oxides/nitride (SiO$_2$, Al$_2$O$_3$ Si$_3$N$_4$) as isolation layer for the HDM. We are mainly relying on self-leveling properties of the polymers and chemical mechanical planarization (CMP) of oxides/nitride to have a flat and smooth surface. Moreover, it is required to be able to process the isolation layer at temperatures lower than the maximum thermal budget of the PNZT thin films to avoid inter diffusion of PNZT and platinum, which might degrade the PNZT properties.

For the reflective top layer, aluminum (Al) is the preferred choice due to its high reflectance ($>75\%$) in the visible spectral range for a minimum film thickness of $\geq20 nm$, irrespective of the underlying isolation layer. Al is selectively deposited by evaporation on top of the flat and smooth isolation layer (right above the active pixel area) using the lift-off process. The choice of Al is not only based on the wavelength range of interest but also on its relatively better oxidation resistance, low residual stress, low surface roughness, adhesion to underlying layer, and ease of processing compared to other potential metals as mirror surface like silver or gold. Furthermore, a protective coating of dielectric material (like TiO$_2$) can be applied to prevent the mirror surface from oxidation and scratches, while maintaining the high reflectance.

\section{Development of a Highly Hysteretic Piezo Material}\label{Material}
\begin{figure}[htbp]\centering
\begin{tikzpicture}
\node at (0,0) {\includegraphics[scale=.5]{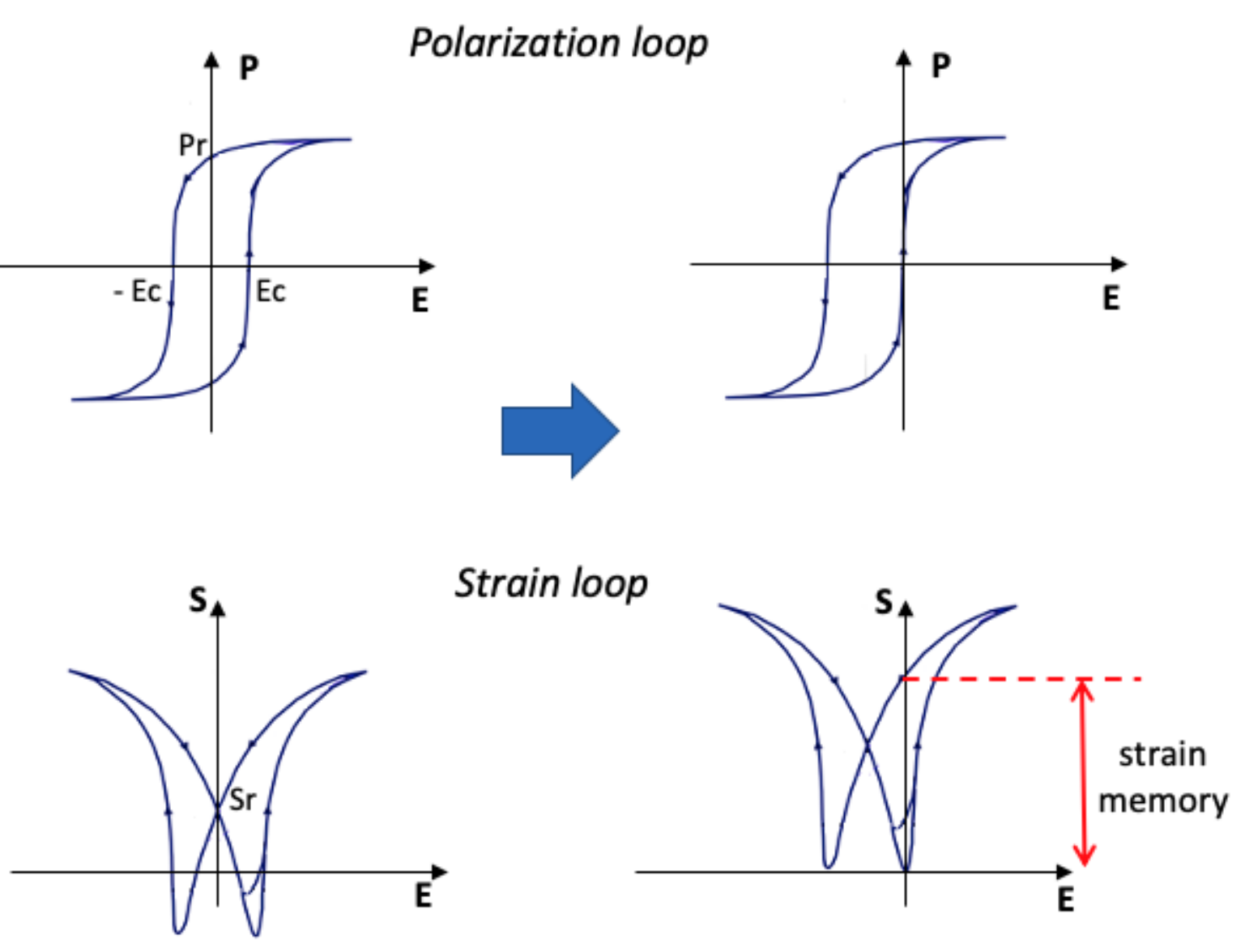}};
\end{tikzpicture}
\caption{Typical polarization and strain loops of high-strain piezoelectric ceramics, such as PZT (left) are compared with those showing strain memory effect (right)} 
\label{fig:PZTmemoryeffect}
\end{figure}

The term strain memory effect is used to indicate the existence, for some piezoelectric materials, of two stable piezoelectric strain states at zero field, that can be accessed through the application of a specific voltage cycle. This corresponds to the display of a shifted Strain - Electric Field (S-E) loop (also known as butterfly loop), which is asymmetric with respect to the piezoelectric strain axis. Figure \ref{fig:PZTmemoryeffect} shows typical P-E and S-E loops of a piezoelectric PZT ceramic. The relation between applied electric field and the induced displacement is highly nonlinear but a negligible remnant deformation is observed after removing the field.

The developed material is based on the well-known family of PZT (the trade name for lead zirconate titanate or PbZr$_{1-x}$Ti$_{x}$O$_3$) with composition close to the so-called morphotropic phase boundary (MPB), which is the Zr/Ti ratio at which a phase transition takes place between a tetragonal and a rhombohedral ferroelectric phase (for $x~0.48$). The MPB compositions are those for which the best piezoelectric parameters and electromechanical coupling factors are found. Other properties are tailored through compositional modification and chemical doping in order to optimize the hysteretic piezoelectric actuators. In particular, we have synthesized a promising candidate for the fabrication of the set-and-forget actuator, namely Pb(Zr$_{0.52}$\Ti$_{0.48}$)$_{0.96}$Nb$_{0.04}$O$_{3}$ with zirconia microparticles, which has been called PNZT. This material exhibits a large longitudinal piezoelectric parameter and stability to fatigue. Furthermore, the strain memory effect has been observed and good mechanical properties have been  obtained due to the addition of $\Zr\OO_{2}$ particles. 

Prior to the development of the HDM, as described in Section \ref{HDM_Concept}, a proof of concept will be manufactured using a single thick ceramic pellet of PNZT. This pressed PNZT pellet will have a diameter of $5 mm$, a thickness between $100$ and $500 \mu m$ and a $5\times 5$ electrode layout.
The final HDM will be build up out of a large stack of PNZT thin films. Apart from the already discussed advantages of the use of thin layers of piezo material, the high homogeneity of these thin films guarantees identical pixel response over the complete mirror surface. In Section \ref{PNZThyspellet} and \ref{PNZThysfilm} respectively, the development of the PNZT pellets and the PNZT thin films is shortly discussed.  

\subsection{PNZT hysteretic pellet}\label{PNZThyspellet}
The piezoelectric pellets are composed by a Ti-rich, Nb-doped PZT (PNZT) matrix with embedded $\Zr\OO_{2}$ micro particles. The presence of Niobium and $\Zr\OO_{2}$ particles in the PNZT matrix influences the piezoelectric response and the mechanical properties by lowering the switching fields and enhancing the piezoelectric coefficient $\dd_{33}$ and the piezoelectric voltage constant $\Gg_{33}$ \cite{Bencan2012, Nguyen2014, Chu2004}. Niobium ($\Nb^{5+}$) as a soft dopant has been selected to increase the piezoelectric coefficients and produce large remnant polarization. Moreover, it has been reported that $\Nb^{5+}$ induces the memory effect by producing a horizontal shift in the S-E loop, making the butterfly loop asymmetric with respect to the piezoelectric strain axis, due to the presence of an internal bias field \cite{Klissurska1997}.

\begin{figure}[htbp]\centering
\begin{tikzpicture}
\node at (-7,0.1) {\includegraphics[scale=.23]{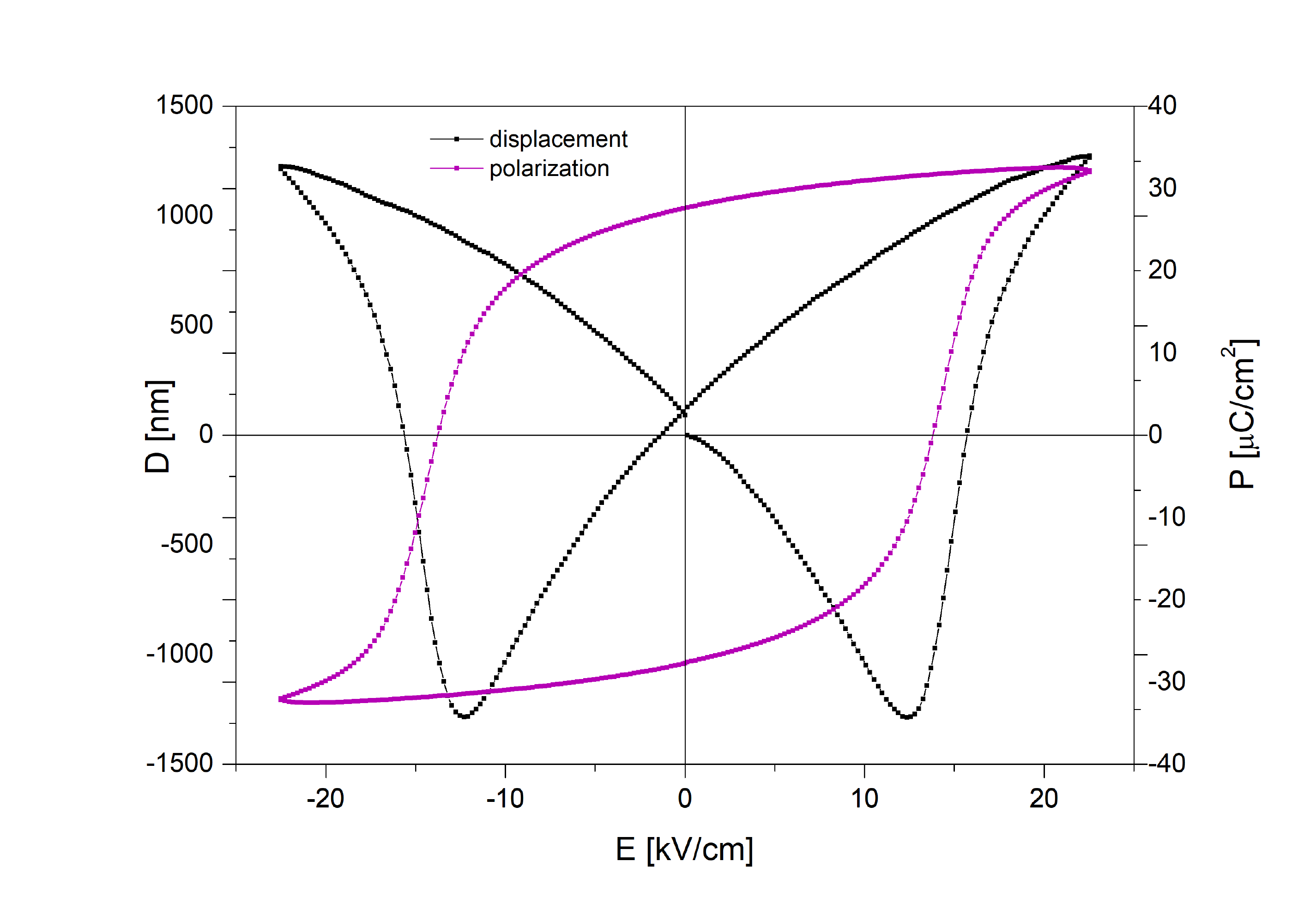}};
\node at ( -0.2,0) {\includegraphics[scale=.38]{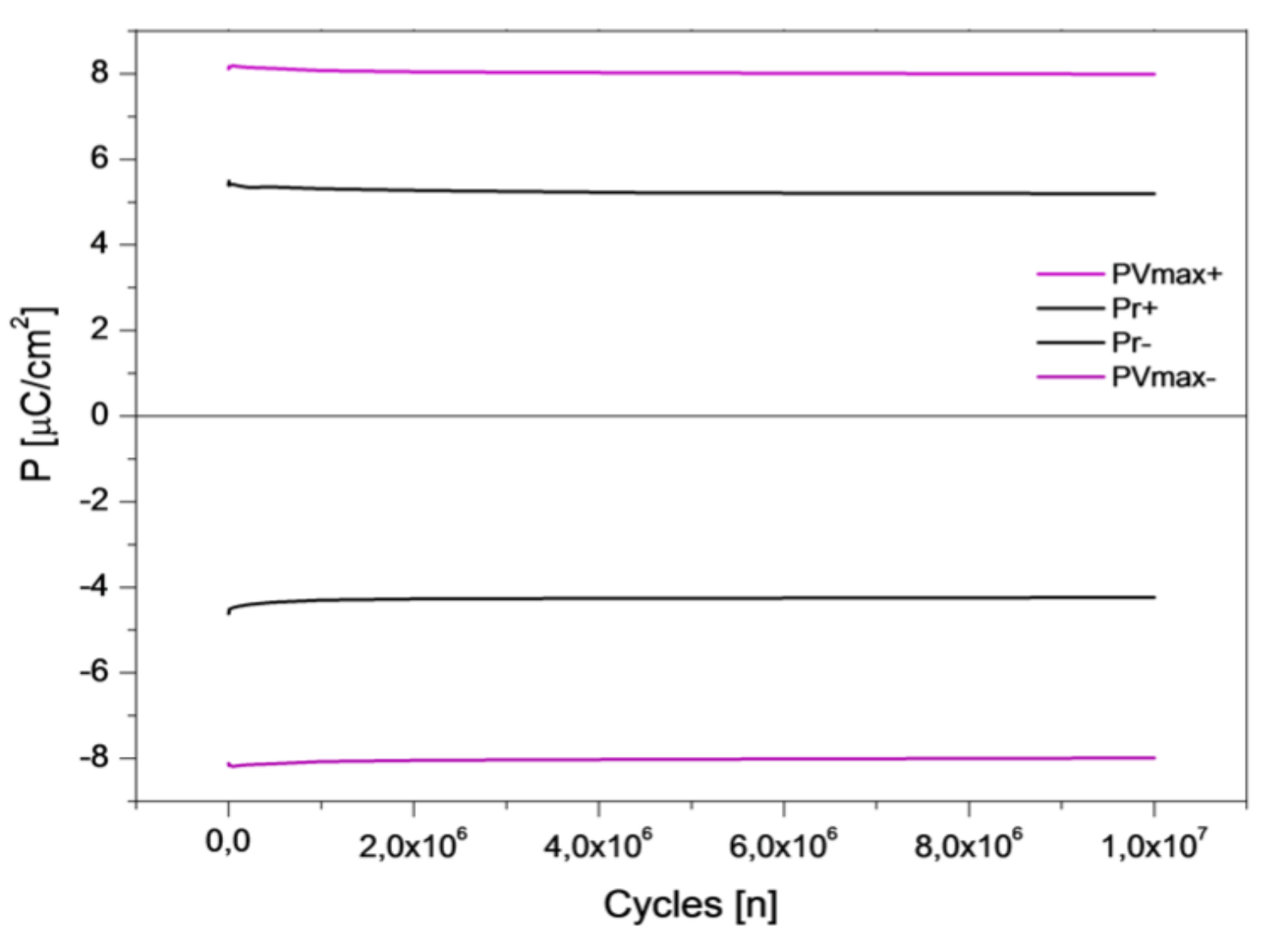}};
\node at (-7,-2.8) {a)};
\node at ( -0.2,-2.8) {b)};
\end{tikzpicture}
\caption{Piezoelectric properties of PNZT hysteretic pellet under high bipolar electric fields: a) Ferroelectric loops and b) Fatigue response.} 
\label{fig:PNZTproperties}
\end{figure}

The conventional mixed oxide route has been used to fabricate PNZT pellets and the precursors are high purity powders of PbO (99.999 \%), $\Zr\OO_{2}$ (99.8 \%), $\Ti\OO_{2}$ (99.6 \%) and $\Nb_{2}\OO_{5}$ (99.99 \%). The precursors were mixed in stoichiometric amounts with 4\% excess PbO to compensate for lead loss during sintering. Afterwards, the pellets were shaped with diameters of $5 mm$ and $10 mm$, and with thickness of $0.5 mm$. Then, the samples were sintered at high temperatures. After the sintering process the samples were poled and the crystal structure and topography was investigated by X-ray Diffraction (XRD) and Scanning Electron Microscopy (SEM). The piezoelectric properties were studied using a state-of-the-art ferroelectric/piezoelectric system from aixACCT. As observed in Figure \ref{fig:PNZTproperties}, the samples exhibit a large piezoelectric deformation, with a maximum displacement of $1.2 \mu m$ at a field of $20 kV/cm$, which corresponds to an effective piezoelectric $\dd_{33}$-coefficient of up to $1200\ \sfrac{\PM}{\V}$. The samples also show good stability to fatigue of at least $10^7$ cycles at high electric field, when the material is fully polarized and its maximum piezoelectric deformation has been reached \cite{Damerio2017}.

\begin{figure}[htbp]\centering
\begin{tikzpicture}
\node at (0,0) {\includegraphics[scale=.5]{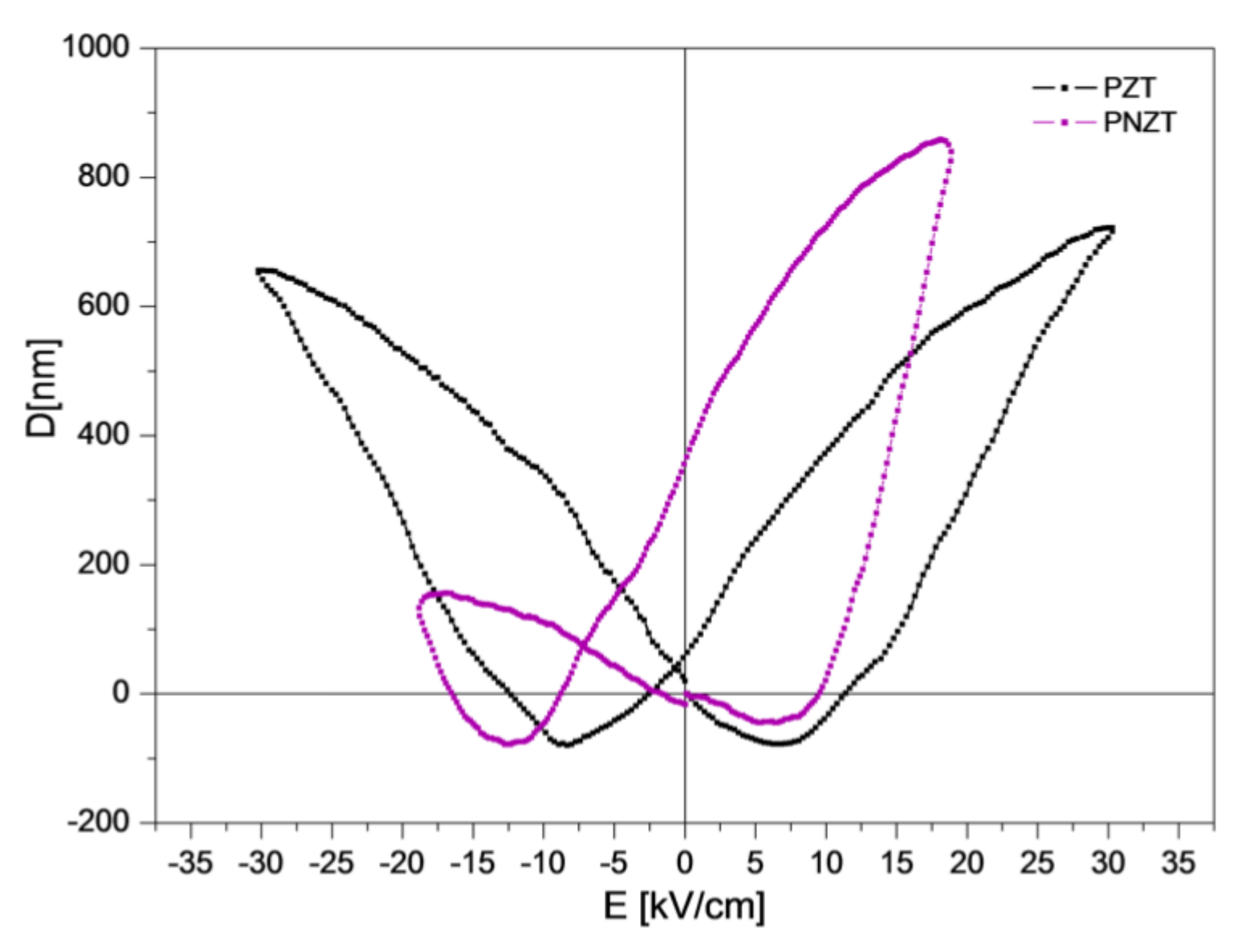}};
\draw[-latex,very thick,green!50!black] (-.5,2) -- (0.45,.2);
\node[green!50!black, text width=3cm,align=center] at (-.5,2.5) {\footnotesize maximum remnant \\ deformation};
\end{tikzpicture}
\caption{Measured strain-electric field (S-E) loops of PZT (black) and PNZT (purple) pellets at $0.5 Hz$. For PNZT, a maximum remnant deformation of approximately $400 nm$ is observed, which is more than $40\%$ of the maximum deformation of the material.} 
\label{fig:PNZTmemoryeffect}
\end{figure}

Figure \ref{fig:PNZTmemoryeffect} shows a comparison between the asymmetric S-E loop of undoped morphotropic PZT and that of the newly developed piezoelectric material PNZT at weak E-fields. For the PNZT sample, a large remnant deformation is present after E-field cycling. The amount of remnant deformation can be controlled by the amplitude of the applied field. We observe $40\%$ remnant deformation \cite{Damerio2017}, which is sufficient for the implementation of multiplexing, allowing for a significant reduction of the device power consumption.

\subsection{Multilayered - PNZT thin films}\label{PNZThysfilm}
For the production of the PNZT thin films, we are developing a chemical solution which enables the deposition of multiple nano layers of the PNZT material by the process of spin coating \cite{VanderVeer2019,VanderVeer2020}. This technique has been chosen because it allows for the preparation of thicker ($1-10 \mu m$) and very homogeneous thin films with large diameter, which facilitates the transition to large-scale industrial applications.

The PNZT chemical solution has the chemical formula $\Pb_{1-z}\Nb_{y}(\Zr_{0.52}\Ti_{0.48})_{1-y}\OO_{3} $ based on ethylene glycol as solvent and bridging ligand. Sols were made with concentrations ranging from $1 M$ to $1.5 M$, lead excesses ($z$) of 0 atom.$ \%$, 10 atom.$ \%$ or 20 atom.$\%$ and niobium doping concentration (y) of 0 atom.$ \%$ or 4 atom.$\%$. The excess of lead was used to compensate for losses due to evaporation during the heat treatments. Additionally, during the synthesis, the influence of the substrate material and the heat treatment of the thin films have been investigated. 

As shown in \cite{VanderVeer2020}, the best piezoelectric properties were obtained from a nine nano layers film (nine spin-coating repetitions) deposited from a $1.5 M$ PNZT sol with a $10\%$ lead excess onto a double-side polished platinized substrate annealed at $650 ^{\circ}C$. The nine-layered film has a thickness of approximately $440 nm$. The ferroelectric and piezoelectric measurements exhibit hysteresis polarization and strain loops of the film at $100 Hz$ up to a field of $800 kV/cm$. The sample has a piezoelectric coefficient $\dd_{33}$ up to $50 \,\sfrac{\PM}{\V}$ and good stability to fatigue of at least $10^7$ cycles \cite{VanderVeer2020}.  

The maximum deformation that can be achieved under the application on an electric field is $0.3\%$ of the total film thickness \cite{VanderVeer2020}, which corresponds to the values found in literature \cite{Balma2014, Klissurska1995}. A slight shift of the ferroelectric strain loop is observed, which indicates the presence of an internal bias field. However, further research is needed to get the desirable memory effect of the film and required maximum deformation for our deformable mirror application. In particular, the multilayered deposition technique demands optimization to increase the homogeneity of the films; a more detailed study of the effects of $\Nb^{5+}$ as a soft dopant in thin films is needed, as these do not need to be the same as in bulk PNZT. Moreover, incorporation of the $\Zr\OO_{2}$ micro particles into the (multilayered) PNZT thin film is still in progress.

\section{Mechanical Modelling and Control of the HDM}\label{Control}
\subsection{Modelling of highly hysteretic piezo actuator}
In literature, the phenomenon of hysteresis has been widely studied and there exist different approaches to describe it mathematically \cite{Preisach1935}.
Despite its mathematical complexity, when it comes to practical applications, the Preisach hysteresis operator has been the preferred mathematical model due to its suitability to describe the hysteresis of different physical phenomena \cite{HASSANI2014209,Fatikow2016}.

To describe the hysteresis present in the relation between the electric field and strain/stress of the HDM piezo actuators’, we have adopted a modified version of the Preisach hysteresis operator, whose weighting function parameter is allowed to have positive and negative values with a particular distribution. This class of Preisach hysteresis operator is able to describe the so-called butterfly hysteresis loops that are experimentally observed in samples of the piezoelectric material used in the HDM as presented in Section \ref{Material}.

\begin{figure}[htbp]\centering
\begin{tikzpicture}
\node at (-4,0) {\includegraphics[scale=.7]{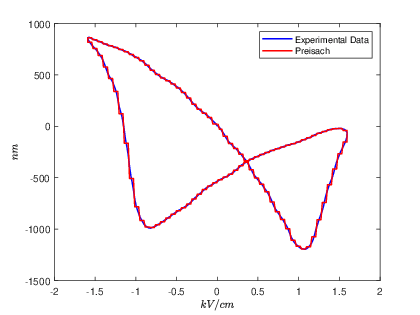}};
\end{tikzpicture}
\caption{Preisach fitting result} 
\label{fig:Preisachfit}
\end{figure}

Figure \ref{fig:Preisachfit} shows the fitting result of the Preisach model parameters to the experimentally obtained PNZT S-E loop. Although the fit is very precise, currently the accuracy of the predicted shape changes over the full range of operation, is typically limited to approximately $90\%$. 

\subsection{Surface shape modelling as input to control}\label{Surfaceshape}
To set each actuator to a desired position for correcting the wavefront aberration, the reflective top layer was modelled by a detailed consideration of pixel shape and position. 

To describe the shape changes of the reflective top layer, we considered the Poisson equation \cite{Morse1953} in polar coordinates, governing the relation between small surface displacements of the thin top layer and surface tension generated by the pixels. Incorporating the HDM's particular arrangement of the electrodes into the solution to Poisson's equation provides the deformation of the reflective top layer at a specific surface point.

To include the mechanical behavior of the actuators, each actuator was modelled as a spring in parallel to a force source. The pressure exerted by the actuators at a specific position on the mirror surface comprises two components, a stiffness and force source over an area. Based on this implementation, the input for the Preisach hysteresis model, used to control the shape of the mirror, can be independently determined.
Using Zernike polynomials as preferred representation for optical wavefront aberrations, the mirror surface was fit to the desired shapes based on selected surface points. To determine the optimal pressure for each actuator, an over-determined set of equations was solved in the least-square sense, minimizing the root-mean-square deviation between actual and desired shape. This information serves as valuable input for the control of the mirror shape as described in Section \ref{Controlstrat}.\\
For a complete description of the surface shape modelling of the HDM, the reader is referred to \cite{PartII}.

\subsection{Control strategy}\label{Controlstrat}
\begin{figure}
    \centering
    \includegraphics[scale=0.5]{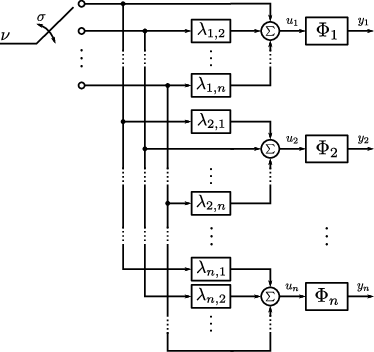}
    \caption{Simplified model of the electrically coupled actuators used in the control strategy of the HDM.}
    \label{fig:simplified_model_control}
\end{figure}

To aid the design of a HDM control strategy, a simplified model of the electrically coupled actuators is considered, where each actuator is modeled by an individual Preisach operator $\Phi_i$, as illustrated in Figure \ref{fig:simplified_model_control}. In this model, two signals are considered: a control signal $\nu$ whose value corresponds to the applied electric field, and a switching signal $\sigma$ whose value indicates the actuator to which the control signal is being applied. In addition, the effects of the electric coupling are captured by the factors $\lambda_{i,j}$, whose values are always less than one. Thus, when the switching signal indicates that the actuator $i$ is selected, the control signal $\nu$ is directly applied to the actuator $i$. The other actuators receive an attenuated version of the control signal $\nu$ scaled by the factors $\lambda_{*,i}$.

\begin{figure}
    \centering
    \subfigure[Actuator 1]{\includegraphics[width=0.32\linewidth]{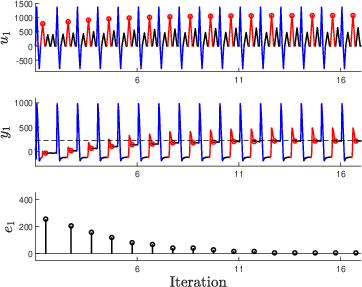}}
    \subfigure[Actuator 2]{\includegraphics[width=0.32\linewidth]{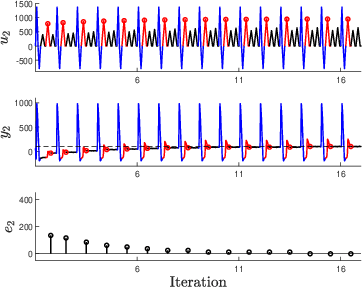}}
    \subfigure[Actuator 3]{\includegraphics[width=0.32\linewidth]{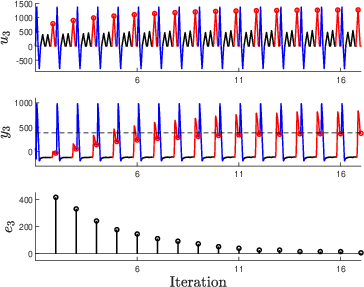}}
    \caption{Simulation of the first $17$ iterations of Algorithm \ref{alg:control_pseudoalgo} for $3$ electrically-coupled identical actuators in HDM. From left to right, each sub-figure corresponds to the closed-loop behaviour of one actuator system that is placed according to the same sequence found in these sub-figures. The corresponding actuator input $u_i$ is at the top with the triangular pulse amplitudes $w_i$ at every iteration marked by a red circle, the corresponding pressure output $y_i$ is in the middle with the remnant pressure at every iteration marked also by a red circle and the reference value $y_i^{\textnormal{ref}}$ indicated by a dashed line, and the corresponding error $e_i$ after every iteration is at the bottom. The blue segments of the inputs and outputs indicate the periods when the reset signal is applied to all actuators and the red segments indicate the periods when the switching signal $\sigma$ selects the corresponding actuator and the control signal $\nu$ is directly applied (i.e. $\sigma=i$ and $u_i=\nu$). It can be observed that when the control input is directly applied to one actuator, the other two receive a $50\%$ attenuated version of it indicated by the black segments of the inputs.}
    \label{fig:control_sim}
\end{figure}

Based on this model, and using the optimal pressure computed for each actuator as a reference (see Section \ref{Surfaceshape}), a control loop is proposed to drive sequentially the actuators of the HDM to their reference value. The strategy to control the complete HDM is inspired by the Iterative Learning Control Methodology and based on the work introduced in \cite{PartIII} which addresses the problem of driving the remnant of a single hysteretic actuator whose behavior can be described by a Preisach operator. In the case of the HDM, the proposed control algorithm is presented in Algorithm \ref{alg:control_pseudoalgo}. This algorithm relies on the possibility to reset all the actuators of the mirror at the same time and the so-called consistency property of hysteresis and the Preisach operator which allows to disregard the effect of the electrical coupling when the input amplitudes are within an allowed range depending on the coupling factors $\lambda_{i,j}$.
Roughly speaking, to set the HDM to a certain configuration, it is considered that the actuators of the HDM can be set to a known initial state and posteriorly an iteration of the control loop can run where the switching signal selects each actuator of the HDM once for a predefined period of time. During this period, the selected actuator receives a control signal consisting of a triangular pulse whose amplitude is computed based on the error between its corresponding reference value and the actual pressure of the actuator at the previous iteration. Correspondingly, due to the effect of the electric coupling, the neighboring actuators receive an attenuated version of the triangular pulse within the same time period. When our proposed iterative control algorithm is applied to each pixel, the effect of this attenuated pulse to the neighboring pixels 
can be avoided 
by {\it apriori} determining 
the actuators' update sequence in the ascending order of magnitude of the triangular pulse to be applied. The iterations of the control loop are repeated until for every single pixel the appropriate pulse amplitude is found and the desired surface shape is achieved. To illustrate this control strategy, a simulation of three coupled actuators described by the model of Figure \ref{fig:simplified_model_control} was performed and the results are presented in Figure \ref{fig:control_sim}. In this simulation, the actuators are driven to strain reference values of $220nm$ for actuator $1$, $100nm$ for actuator $2$ and $380nm$ for actuator $3$. All coupling factors $\lambda_{i,j}$ are $0.5$. The initial pulses amplitudes for the three actuators were set to $800V$ and the gain used for the pulse amplitude update at every iteration was $\gamma=0.28$. 

\begin{algorithm}[H]
    \SetAlgoLined
    \SetKwInput{kwInit}{Init}
    \KwIn{Number of actuators: $n_a$, pressure reference vector: $y^{\textnormal{ref}}$, error threshold: $e^{\text{threshold}}$, update gain: $\gamma$}
    \kwInit{set amplitudes: $w_i=0 \textnormal{ for all } i\in\{1\dots n_a\}$}
    \While{$\textnormal{not}(e_i<e^{\textnormal{threshold}} \textnormal{ for all } i\in\{1\dots n_a\})$}{
        reset all actuators \\
        obtain indexes of ascending sorted amplitudes:\\
        \qquad\qquad $idxs$ = index\_ascending\_sorted(vector $= w$)\\
        \For{$j = idxs$}{
            set switching signal: $\sigma = j$\\
            apply input: $\nu$ = triangle\_pulse(amplitude = $w_j$)\\
            compute error: $e_j = y_j-y_j^{\textnormal{ref}}$ \\
            update amplitude: $w_j = w_j + \gamma \cdot e_j$
        }
    }
    \caption{HDM Control algorithm}
    \label{alg:control_pseudoalgo}
\end{algorithm}

\section{Discussion}\label{Discussion}
The extreme DM requirements, applicable to the direct detection of exoplanets, in combination with the limitations imposed by both the space craft and the harsh space environment, form a substantial challenge for the hardware development. These demand the use of large stacks of the PNZT thin films to be able to meet functional specs of low voltage and high pixel density and to optimize performance.\\
When we limit the allowed deformation to $0.1\%$ of its total thickness, for $10 \mu m$ thick thin films a stack of 250 layers would be required to reach a maximum deformation of $1 \mu m$ (considering $40\%$ remnant deformation).
To limit the number of independent piezo layers, an important focus of the PNZT thin film development will be on the realization of relatively thick PNZT thin films. Another approach we are currently investigating is the development of so called "composite PNZT" with a mixture of chemical solution with fillers such as PNZT micro particles. The aim of this study is to produce $10-100 {\mu} m$ thick PNZT layers with still sufficient homogeneity.
Apart from this we are working on automation of parts of the process. This not only makes the production of large amounts of thin films possible but also improves the process control, which is critical for the quality of the final product. 

The PNZT actuator shows similarity to the PI-rest actuator \cite{PIrest} recently developed by Physik Instrumente (PI) GmbH \& Co. Remnant deformation of this piezo actuator can be controlled by changing the polarization state of the material. 
PI-rest is manufactured using multilayer piezo actuators and its active material is soft PZT ceramics. Soft PZT ceramics are characterized by a large piezoelectric constant and good electromechanical coupling factors but also by low mechanical quality factors, easy depolarization and poor linearity \cite{Holterman2013, Lee2006}. Although the PNZT actuator is fabricated using Nb as a soft dopant, the produced PNZT material with $\Zr\OO_{2}$ particles exhibits good mechanical properties, linearity, and good resistance to fatigue and depolarization. Nevertheless, the most distinguish characteristics of PNZT actuators are: 1) fabrication of the device with high actuator number (pixels), which reduces the power consumption and simplifies the electronics and control of the system; and 2) the development of thin films (in addition to pellets) for large scale surface actuators and reduction of the input voltage, that also reduces the overall fabrication cost of the deformable mirror. 

With our development we focus on space based DMs where small stroke but highly accurate corrections for static (imperfections) and slowly moving (thermal) aberrations are required. We strongly believe however that the concept has wide spin off possibilities in the high tech industry like chip manufacturing, high intensity lasers and microscopy. For these applications we are not only considering the application as deformable mirror but also as deformable surface or as single set and forget actuator. As these applications are typically less restrictive in operational conditions (e.g., applied voltage, vacuum conditions, stroke) also the option of PNZT hybrids or even pellets can be considered here. The valuable knowledge gained in this study can directly be put to use when considering these other applications.

\section{Conclusions}\label{Conclusions}
We have introduced the HDM concept and provided an overview of the critical technology steps for the realization of the hardware. The concept, to produce very high pixel numbers in a dense configuration by the application of a simple electrode layout, is supported by FEA of the electric field propagation and deformation of a single layer of standard PZT material. The control strategy accounts for the observed deformation of the neighbouring pixels and brings the mirror to its final shape in an iterative process. Both the pixel addressability and the use of TDM rely on the development of piezo materials with large remnant deformation. Efforts have been devoted to develop such a material and here we show our progress in this direction presenting a PZT-based composite ceramic that exhibits $40\%$ remnant deformation, which is sufficient to serve as set-and-forget actuator.

Our aim is to develop a pellet based demonstrator. This $5\times 5$ pixel DM will serve as proof of principle and should validate the concept of single pixel addressing by TDM in combination with the simple electrode layout. It should also verify that the set-and-forget nature of the PNZT results in stable shape configurations without the need for constant actuation or a high update frequency and the ability to shape the mirror by application of the developed control strategy. In parallel we will work on the further development of the material, the control and the fabrication steps involved with the production and operation of a final HDM.

\section*{Funding}
This work was co-funded through a Marie Skłodowska-Curie COFUND (DSSC 754315), a NSO/NWO PIPP grant and a FOM/f Fellowship of the Dutch Research Council (NWO).

\section*{Acknowledgments}
The authors would like to thank Wouter van de Beek for his contribution to the development of the HDM Preisach operator and J. Baas and H. Bonder for their technical support.

\section*{Disclosures}
The authors declare no conflicts of interest.

\bibliography{Biblio} 

\end{document}